\documentclass[prd,twocolumn,aps]{revtex4-1}

\usepackage{bm,graphicx,textcomp,amssymb,amsmath,dcolumn,hyperref,color}

\newcommand{\ud}{\mathrm{d}}
\newcommand{\be}{\begin{equation}}
\newcommand{\ee}{\end{equation}}
\newcommand{\LCm}{{\scriptscriptstyle -}} 
\newcommand{\LCp}{{\scriptscriptstyle +}}

\newcommand{\LCperp}{{\scriptscriptstyle \perp}}

\newcommand{\braket}[2]{\left<#1|#2\right>}

\newcommand{\f}[1]{\mbox{\boldmath$#1$}}

\newcommand{\bea}{\begin{eqnarray}}
\newcommand{\ea}{\end{eqnarray}}
\newcommand{\eea}{\end{eqnarray}}

\usepackage{siunitx}
\newcommand{\diff}{\mathrm{d}}
\newcommand{\pathd}{\mathcal{D}}

\newcommand{\mat}[1]{\boldsymbol{\mathrm{#1}}}
\DeclareMathOperator{\tr}{tr}

\begin{document}

\title{Sauter-Schwinger pair creation dynamically assisted by a plane wave}

\author{Greger Torgrimsson$^{1,2,3}$}
\author{Christian Schneider$^{1}$}
\author{Ralf Sch\"utzhold$^{1}$}

\affiliation{$^1$Fakult\"at f\"ur Physik, Universit\"at Duisburg-Essen, 
Lotharstra{\ss}e 1, Duisburg 47048, Germany,}

\affiliation{$^2$Friedrich-Schiller-Universit\"at Jena, Max-Wien-Platz 1, 
07743 Jena, Germany,} 

\affiliation{$^3$Helmholtz Institute Jena, Fr\"obelstieg 3,
07743 Jena, Germany.}

\date{\today}

\begin{abstract}
We study electron-positron pair creation by a strong and constant electric 
field superimposed with a weaker transversal plane wave which is incident 
perpendicularly (or under some angle). 
Comparing the fully non-perturbative approach based on the world-line 
instanton method with a perturbative expansion into powers of the strength 
of the weaker plane wave, we find good agreement -- provided that the 
latter is carried out to sufficiently high orders. 
As usual for the dynamically assisted Sauter-Schwinger effect, the additional 
plane wave induces an exponential enhancement of the pair-creation probability 
if the combined Keldysh parameter exceeds a certain threshold. 
\end{abstract}

%\pacs{}

\maketitle

%%%%%%%%%%%%%%%%%%%%%%%%%%%%%%%%%%%%%%%%%%%%%%%%%%%%%%%%%%%%%%%%%%%%%%%%%%%%%%%
\section{Introduction}
%%%%%%%%%%%%%%%%%%%%%%%%%%%%%%%%%%%%%%%%%%%%%%%%%%%%%%%%%%%%%%%%%%%%%%%%%%%%%%%

As one of the most striking consequences of quantum field theory, 
extreme external conditions can tear apart quantum vacuum fluctuations 
and thereby create real particles. 
Already in 1939, Schr\"odinger predicted that the rapid expansion of the 
Universe could induce such a process \cite{Schroedinger}.
As another example, the strong gravitational field around a black hole can 
tear apart quantum vacuum fluctuations leading to Hawking radiation, 
i.e., black-hole evaporation \cite{Hawking74,Hawking75}. 
In analogy to the gravitational force, a strong electric field can have 
a similar effect and create electron-positron pairs out of the quantum 
vacuum -- the Sauter-Schwinger effect 
\cite{Sauter1931,Sauter1932,Heisenberg1936,Schwinger}.
For a constant electric field $E$, the pair creation probability 
(per unit time and volume) scales as ($\hbar=c=1$) 
\bea
\label{probability}
P_{e^+e^-}
\sim 
\exp\left\{-\pi\,\frac{m^2}{qE}\right\} 
\,,
\ea
where $q$ and $m$ are the elementary charge and the mass of the 
positron/electron, respectively.  

Unfortunately, this fundamental prediction of quantum field theory has 
not been directly verfied experimentally yet because the required field 
strength is very large. 
This motivates the quest for ways to enhance the pair-creation probability
or, equivalently, to lower the required field strength.
One option is the dynamically assisted Sauter-Schwinger 
effect~\cite{Dynamically-assisted}, where the pair-creation probability 
is strongly enhanced by adding 
a weaker time-dependent field to the strong field $E$. 
So far, most studies of this enhancement mechanism have been devoted to 
purely time-dependent fields \cite{footnote}. 

As a step towards a more relistic field configuration, we consider a 
propagating plane wave superimposed to the constant field $E$ in the 
following. 
Plane waves propagating {\em parallel} to the strong electric field 
were already considered 
in~\cite{Narozhnyi:1976zs,Fradkin:1991zq,Gitman:1996wm}, for example. 
It was found that such transversal plane waves waves do not enhance 
the pair creation probability~\cite{Narozhnyi:1976zs}. 
Further, for longitudinal parallel waves, $E_z(t+z)$, 
the pair creation probability is given by the locally 
constant field approximation~\cite{Tomaras:2000ag,Tomaras:2001vs,
Ilderton:2014mla}, which implies that the enhancement is 
comparably small. 
Both results can be understood by considering a Lorentz boost along the 
direction of the strong field which leaves the strong field invariant but 
reduces the frequency of the plane wave, see also~\cite{Ilderton:2015qda}. 
In the transversal case, the field strength of the plane wave is reduced 
as well while the longitudinal wave retains its field strength.  

In contrast to the {\em parallel} scenarios discussed 
in~\cite{Narozhnyi:1976zs,Fradkin:1991zq,Gitman:1996wm,Tomaras:2000ag,
Tomaras:2001vs,Ilderton:2014mla}, we consider 
the case of a transversal plane wave propagating {\em perpendicular} to the 
strong field $E$
\bea
\label{profile}
\f{E}(t,x)=E\f{e}_z+\varepsilon E\cos(\Omega[t-x])\f{e}_z
\,, 
\ea
corresponding to the vector potential (in temporal gauge) 
$A_z(t,x)=Et+\varepsilon E\sin(\Omega[t-x])/\Omega$. 

This scenario has several advantages:
Since such a transversal wave cannot create electron-positron pairs on its 
own (due to a similar Lorentz boost argument as above), pair creation can 
only occur in cooperation with the strong field $E$, which retains the 
non-perturbative character of this effect. 
Furthermore, the above profile~\eqref{profile} represents a vacuum solution 
to the Maxwell equations and could be a reasonable approximation for an 
experimental set-up where $E$ represents the focus of an optical laser while 
the plane wave is generated by an x-ray free-electron laser (XFEL).  
Finally, we found that this scenario~\eqref{profile} yields the maximum 
enhancement of the pair creation probability. 
Other profiles, polarizations and propagation directions will be discussed 
below in Sec.~\ref{Other directions} and appendix~\ref{Appendix}.  
Note that this profile~\eqref{profile} was already considered in 
\cite{Catalysis}
using first-order perturbation theory in $\varepsilon$, whereas we are 
going to consider higher orders as well as a fully non-perturbative 
approach. 

%%%%%%%%%%%%%%%%%%%%%%%%%%%%%%%%%%%%%%%%%%%%%%%%%%%%%%%%%%%%%%%%%%%%%%%%%%%%%%%
\section{Perturbative approach}
%%%%%%%%%%%%%%%%%%%%%%%%%%%%%%%%%%%%%%%%%%%%%%%%%%%%%%%%%%%%%%%%%%%%%%%%%%%%%%%

At first, we employ a perturbative expansion of the total pair creation 
probability $P_{e^+e^-}$ in powers of the relative strength $\varepsilon$ 
of the plane wave, which is supposed to be small $\varepsilon\ll1$
\bea
\label{expansion}
P_{e^+e^-}=\sum\limits_{N=0}^{\infty}\varepsilon^N P_N 
\,,
\ea
where the contributions $P_N$ can be derived via the world-line formalism, 
for an introduction see~\cite{Schubert:2001he,SchubertLectureNotes} 
and references therein. 
The zeroth order $N=0$ reproduces the original Sauter-Schwinger effect in 
Eq.~\eqref{probability} and odd orders vanish in this situation 
(but not always \cite{example}). 

The lowest-order term $N=2$ corresponding to the one-photon contribution
has already been calculated in \cite{Catalysis}.
Deriving the exponential dependence for the higher-order terms, it turns out 
that the exponent for two photons ($N=4$) with frequency $\Omega$ is the same 
as that for a single photon ($N=2$) with twice the frequency $2\Omega$, 
and so one for more photons (see appendix~\ref{Appendix}).

Consequently, we find 
\bea
\label{P-exp}
P_N\sim\exp\left\{
-\frac{2m^2_\perp}{qE}
\left(\arccos\Sigma-\Sigma\sqrt{1-\Sigma^2}\right)
\right\} 
\,,
\ea
where the function of $\Sigma$ in the exponent 
is already known from~\cite{non-perturbative-versus-perturbative}. 
The effective mass $m_\perp=\sqrt{m^2+(N\Omega/4)^2}$ reflects momentum 
conservation in $x$-direction, where the momentum $N\Omega/2$ of the $N/2$ 
photons has to be transferred to the electron-positron pair.   
As a result, the effective mass $m_\perp$ is higher than the original mass $m$ 
and hence the pair creation probability is lower than in the case of a purely 
time-dependent field.
Finally, $\Sigma$ describes the relative contribution of the energy of the 
$N/2$ photons in comparison to the effective mass gap $2m_\perp$ 
\bea
\Sigma=\frac{N\Omega/2}{2m_\perp}=\frac{N\Omega}{4\sqrt{m^2+(N\Omega/4)^2}}
\,.
\ea
In the limit of $\Omega\downarrow0$, i.e., $\Sigma\downarrow0$, 
where $m_\perp\downarrow m$, we recover Eq.~\eqref{probability},
as expected. 

%%%%%%%%%%%%%%%%%%%%%%%%%%%%%%%%%%%%%%%%%%%%%%%%%%%%%%%%%%%%%%%%%%%%%%%%%%%%%%%
\subsection{Dominant Order}
%%%%%%%%%%%%%%%%%%%%%%%%%%%%%%%%%%%%%%%%%%%%%%%%%%%%%%%%%%%%%%%%%%%%%%%%%%%%%%%

Inspecting the terms in the sum~\eqref{expansion} we find that the prefactors  
$\varepsilon^N$ decrease as $N$ increases (due to $\varepsilon\ll1$) while the 
exponentials in $P_N$ grow according to Eq.~\eqref{P-exp}.
As a result, there could be a dominant order $N^*$ which yields the maximum 
contribution to the sum~\eqref{expansion}.
In order to study this question, we approximately treat $N$ as a continuous 
variable and apply the saddle-point method to the term $\varepsilon^NP_N$, 
i.e., 
\be
\frac{d}{dN}
\left\{
-N|\ln\varepsilon|
-\frac{2m^2_\perp}{qE}
\left(\arccos\Sigma-\Sigma\sqrt{1-\Sigma^2}\right)
\right\} 
=0
\;.
\ee
This yields the dominant order $N_*$ as solution of the transcendental 
equation 
\bea
\label{transcendental}
1-\frac{N_*\Omega}{4m}\,\arctan\frac{4m}{N_*\Omega}
=
\frac{qE}{m\Omega}\,|\ln\varepsilon|
=
\frac{\gamma_{\rm crit}}{\gamma}
\,,
\ea
where we have introduced the (combined) Keldysh parameter 
$\gamma=m\Omega/(qE)$ and its threshold $\gamma_{\rm crit}=|\ln\varepsilon|$.  
We only obtain real solutions $N_*$ if the right-hand side is less than unity, 
i.e., if $\gamma$ exceeds the threshold $\gamma_{\rm crit}$.  
At thershold, $\gamma=\gamma_{\rm crit}$, we find $N_*=0$ which implies the 
original Schwinger result~\eqref{probability}.
For $\gamma>\gamma_{\rm crit}$ and $\Omega\ll m$, however, the dominant order 
$N_*$ can be quite large (which justifies the continuum approximation). 
For example, for $\gamma=3\gamma_{\rm crit}$, the dominant order $N_*$ 
scales as $N_*\sim m^2/(qE|\ln\varepsilon|)$ which can be a large number 
for electric fields $E$ well below the Schwinger limit $E_S=m^2/q$. 

In the limit $\gamma/\gamma_{\rm crit}\gg1$, we may approximate the solution 
of the transcendental equation~\eqref{transcendental} via 
\bea
N_*(\gamma\gg\gamma_{\rm crit})\approx\frac{4m}{\Omega}\,
\sqrt{\frac{\gamma}{3\gamma_{\rm crit}}} 
\,,
\ea
which will also be a large number unless the frequency $\Omega$ far exceeds 
the electron mass $m$. 
Inserting this approximate solution for the dominant order $N_*$ back into 
the exponent~\eqref{P-exp}, we find 
\bea
\label{Plargetildegamma}
P_{e^+e^-}
\sim 
\exp\left\{-8\frac{m^2}{qE}\,\sqrt{\frac{\gamma_{\rm crit}}{3\gamma}}\right\} 
\,.
\ea
In contrast to the dynamically assisted Sauter-Schwinger effect with a purely 
time-dependent field, we see that the exponent still crucially depends on the 
strong field $E$, which demonstrates the non-perturbative character 
of this effect even for $\gamma/\gamma_{\rm crit}\gg1$. 
As mentioned in the Introduction, this is a consequence of the fact that a 
plane wave alone cannot create electron-positron pairs out of the vacuum. 

%%%%%%%%%%%%%%%%%%%%%%%%%%%%%%%%%%%%%%%%%%%%%%%%%%%%%%%%%%%%%%%%%%%%%%%%%%%%%%%
\subsection{Improved Approximation}
%%%%%%%%%%%%%%%%%%%%%%%%%%%%%%%%%%%%%%%%%%%%%%%%%%%%%%%%%%%%%%%%%%%%%%%%%%%%%%%

In the following, we try to improve the accuracy of the approximation outlined
in the previous subsection.
The above estimate of the leading order $N_*$ was based on the competition 
between the factor $\varepsilon^N$ and the exponent~\eqref{P-exp}.  
However, the pre-factor in front of this exponent will also depend on $N$
and thereby slightly modify the scaling with $N$.
Thus, in order to improve our approximation, we make an educated guess for 
the scaling of that pre-factor with $N$.  
Obviously, each additional power of $\varepsilon$ must be accompanied by a 
factor of $qE$ since this governs the coupling to the fermionic field. 
Recalling the structure of the QED interaction (vertex) term 
$\bar\psi qA_\mu\gamma^\mu\psi$, it seems quite reasonable to suppose a 
scaling with $\varepsilon qE/\Omega$ since $A_\mu\propto E/\Omega$.
Finally, in view of dimensionality arguments, we arrive at the following 
rough estimate for the scaling of the pre-factor 
%
%\bea
%P_N
%&\sim&
%\left(\varepsilon\,\frac{qE}{m\Omega}\times{\rm const}\right)^N
%\times
%\nn
%&&
%\times
%\exp\left\{
%-\frac{2m^2_\perp}{qE}
%\left(\arccos\Sigma-\Sigma\sqrt{1-\Sigma^2}\right)
%\right\} 
%\,,
%\ea
%
\be
\begin{split}
\varepsilon^N
P_N
\sim&
\left(
\varepsilon\,
\frac{qE}{m\Omega}\times{\rm const}\right)^N
\times \\
&\times
\exp\left\{
-\frac{2m^2_\perp}{qE}
\left(\arccos\Sigma-\Sigma\sqrt{1-\Sigma^2}\right)
\right\} 
\,,
\end{split}
\ee
where $\rm const$ is a (so far undetermined) constant -- or, more precisely, 
a factor which should only weakly depend on the involved parameters.

\begin{figure}
\includegraphics[width=\linewidth]{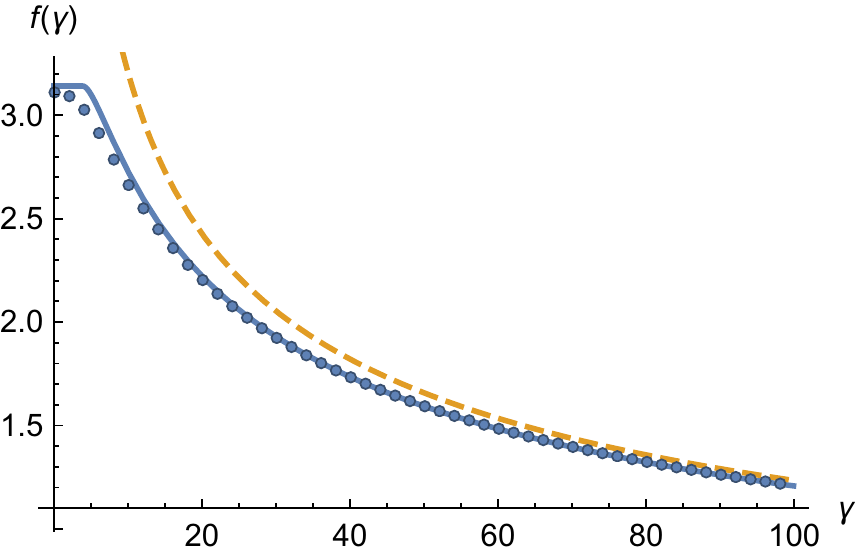}
\includegraphics[width=\linewidth]{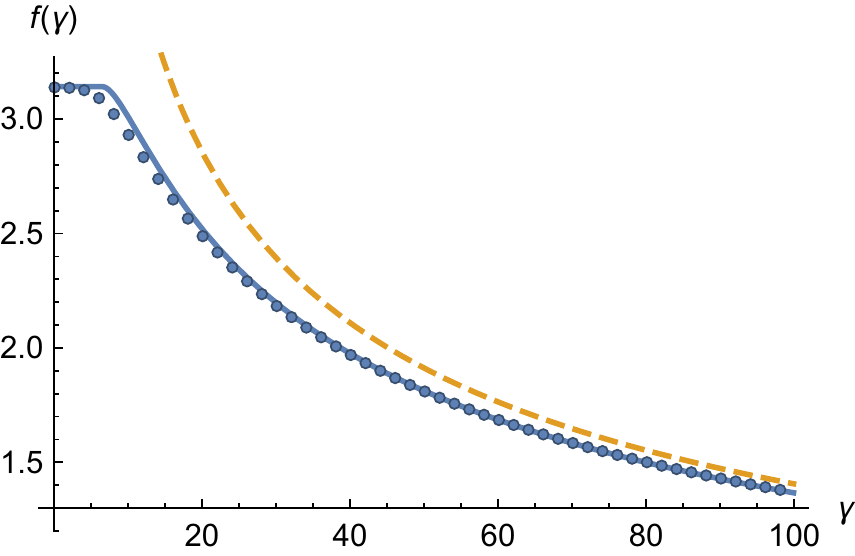}
\caption{
(Color online)
Plot of the exponent of the pair-creation probability $P_{e^\LCp e^\LCm}$ 
as a function of $\gamma$ for perpendicular incidence and parallel 
polarization~\eqref{profile} 
with
$\varepsilon=10^{-2}$ (top) and  
$\varepsilon=10^{-3}$ 
(bottom). 
The exponent has been multiplied by $qE/m^2$, 
i.e., the plot shows $f(\gamma)$ such that 
$P_{e^\LCp e^\LCm}\sim\exp\{-f(\gamma)m^2/[qE]\}$. 
The circles represent the numerical world-line instanton results from 
Sec.~\ref{World-line instantons} and the dashed curve corresponds to 
the large-$\gamma$ approximation in~\eqref{Plargetildegamma}.
The solid curve shows the result of our improved analytical 
approximation obtained by inserting the dominant order $N_*$ from 
Eq.~\eqref{TranscendentalSaddle} into Eq.~\eqref{P-exp}. 
The constant factor in Eq.~\eqref{TranscendentalSaddle} has been chosen 
in order to match the world-line instanton results, which 
gives 
a factor of $8$ for $\varepsilon=10^{-2}$ (top) and 
a factor of $9.5$ for $\varepsilon=10^{-3}$ 
(bottom). 
With these values, we observe good agreement between our improved analytical 
approximation and the numerical world-line instanton results.}
\label{StrongEWeakPWpara-fig}
\end{figure}

\begin{figure}
\includegraphics[width=\linewidth]{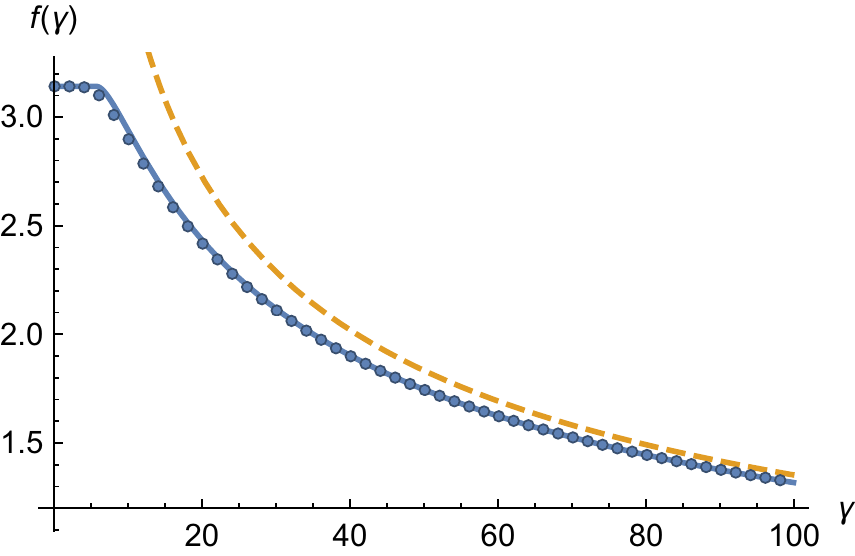}
\includegraphics[width=\linewidth]{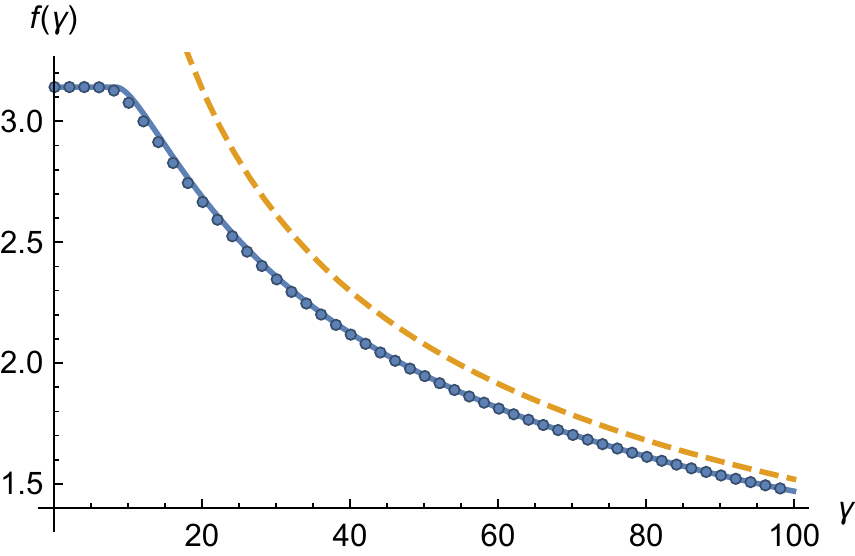}
\caption{(Color online) Same as Fig.~\ref{StrongEWeakPWpara-fig}, 
but for perpendicular polarization $\f{e}_P^\perp$ as discussed 
in Sec.~\ref{Other directions}. 
Again, the constant factor in Eq.~\eqref{TranscendentalSaddle} 
is obtained by fitting 
and gives 
$1.9$ for $\varepsilon=10^{-2}$ (top) and 
$2$ for $\varepsilon=10^{-3}$ (bottom). 
We observe that perpendicular polarization yields a lower 
pair-creation probability $P_{e^\LCp e^\LCm}$.}
\label{StrongEWeakPWperp-fig}
\end{figure}

The saddle point of the above expression gives a slightly shifted dominant 
order $N_*$ as solution of the transcendental equation with a modified 
right-hand side 
%
%\bea
%1-\frac{N_*\Omega}{4m}\,\arctan\frac{4m}{N_*\Omega}=
%\frac{qE}{m\Omega}\,\ln
%\left(\varepsilon\,\frac{qE}{m\Omega}\times{\rm const}\right)
%\,.
%\ea
%
\be
\label{TranscendentalSaddle}
1-\frac{N_*\Omega}{4m}\,\arctan\frac{4m}{N_*\Omega}=
\frac{qE}{m\Omega}\,\left|\ln
\left(\varepsilon\,\frac{qE}{m\Omega}\times{\rm const}\right)\right|
\,.
\ee
Comparison with fully non-perturbative results obtained with the world-line 
instanton method as described in Sec.~\ref{World-line instantons} shows 
that this modification is indeed an improvement of our approximation and 
leads to good agreement, see Figs~\ref{StrongEWeakPWpara-fig} 
and~\ref{StrongEWeakPWperp-fig}. 

%

%%%%%%%%%%%%%%%%%%%%%%%%%%%%%%%%%%%%%%%%%%%%%%%%%%%%%%%%%%%%%%%%%%%%%%%%%%%%%%%
\section{Other directions}\label{Other directions}
%%%%%%%%%%%%%%%%%%%%%%%%%%%%%%%%%%%%%%%%%%%%%%%%%%%%%%%%%%%%%%%%%%%%%%%%%%%%%%%

So far, we have considered the case of perpendicular incidence and a plane wave with the electric field component parallel to
the strong field~\eqref{profile}, which yields the 
maximum enhancement.
Now let us briefly discuss more general angles 
\bea
\label{profile-directions}
\f{E}(t,\f{r})=E\f{e}_z+\varepsilon E\cos(\Omega t-\f{K}\!\cdot\!\f{r})
\f{e}_P
\,, 
\ea
where the corresponding vector potential reads $A_0=0$ and 
$\f{A}(t,\f{r})=
Et\f{e}_z+\varepsilon E\f{e}_P\sin(\Omega t-\f{K}\!\cdot\!\f{r})/\Omega$. 
Without loss of generality, we may set 
$\f{K}=K_\|\f{e}_z+K_\perp\f{e}_x$ where $K_\perp\geq0$. 
The polarization vector $\f{e}_P$ obeys $\f{K}\!\cdot\!\f{e}_P=0$, 
for example $\f{e}_P^\perp=\f{e}_y$.

Inserting this more general field profile~\eqref{profile-directions}, 
we obtain formally the same results as in the previous section with 
$\Omega$ being replaced by $K_\perp$.  
Since the enhancement of the pair-creation probability is monotonic 
in $\Omega$ (i.e., now $K_\perp$), we see that the perpendicular case 
with $K_\|=0$ and $K_\perp=\Omega$ is indeed optimal, see also 
\cite{Catalysis}. 

Note that, while the exponents do not depend on the polarization vector 
$\f{e}_P$, the pre-factors do depend on $\f{e}_P$, cf.~\cite{Catalysis}, 
which can also generate a slight polarization dependence 
of the dominant order $N_*$ 
via the constant factor in~\eqref{TranscendentalSaddle}. 
This is consistent with the results of the next Section, which show that 
the world-line instantons and their actions do also depend on the 
polarization vector $\f{e}_P$. 

%%%%%%%%%%%%%%%%%%%%%%%%%%%%%%%%%%%%%%%%%%%%%%%%%%%%%%%%%%%%%%%%%%%%%%%%%%%%%%%
\section{World-line instantons}\label{World-line instantons}
%%%%%%%%%%%%%%%%%%%%%%%%%%%%%%%%%%%%%%%%%%%%%%%%%%%%%%%%%%%%%%%%%%%%%%%%%%%%%%%

We can express the probability for pair creation using the vacuum
persistence amplitude $\braket{0_\text{out}}{0_\text{in}}$
and thus the effective action $\Gamma$ 
with $\braket{0_\text{out}}{0_\text{in}}=e^{i\Gamma}$
\begin{equation}
  P_{e^+e^-}=1 - \left|\braket{0_\text{out}}{0_\text{in}}\right|^2
  = 1 - e^{-2 \Im \Gamma} \approx 2\Im\Gamma.
\end{equation}
The world-line instanton method is a semiclassical evaluation of the
world-line path integral for the euclidean effective
action~\cite{Affleck1982, Dunne2005, Dunne2006a} (euclidean because we
replace time by imaginary time $x_4=i t$, it is related to the
Minkowskian quantity by $\Gamma = i\Gamma^{\mathrm{E}}$~\cite{Dunne2006a}),
\begin{multline}
\label{eq:WL-path-integral}
\Gamma^\mathrm{E} = \int_0^\infty\frac{\diff{T}}{T} e^{-\frac{m^2}{2} T}
\int\pathd x(\tau) 
\\
\Phi[x]\exp\left[-\int_0^T\!\!\!\diff{\tau}
\left(\frac{\dot{x}^2}{2} + iq A^\mathrm{E}\cdot \dot{x}
  % -\frac{1}{4}\sigma_{\mu\nu}iqF^\mathrm{E}_{\mu\nu}]
\right)\right],
\end{multline}
where the paths $x(\tau)$ are (in general 4-dimensional) closed
trajectories parametrized by proper time $\tau$ with period
$T$, $A^\mathrm{E}_\mu(x(\tau))$ is the euclidean four-potential
evaluated on the trajectory and
\begin{equation}
  \Phi[x] = \frac{1}{2}\tr_\Gamma \mathcal{P} e^{\frac{1}{4}\int_0^T\diff{\tau}\
  \sigma_{\mu\nu}iqF^\mathrm{E}_{\mu\nu}(x(\tau))},
\end{equation}
is the spin factor with
$\sigma_{\mu\nu}=\frac{1}{2}\left[\gamma_\mu, \gamma_\nu\right]$,
$\tr_\Gamma$ denoting the trace over spinor indices and
$\mathcal{P}e^{\dots}$ the path ordered exponential. In this section
we will only work with euclidean quantities, so we omit the
superscript ${}^\mathrm{E}$ for brevity.

A saddle point evaluation for both the $T$- and the path integral
consists of finding periodic solutions that satisfy the Euler-Lagrange
equations corresponding to the exponent
in~\eqref{eq:WL-path-integral}, that is,
\begin{equation}
  \label{eq:instanton-eom}
  m\ddot{x}_\mu(\tau) = a q\, iF_{\mu\nu}(x(\tau))\dot{x}_\nu(\tau),
  \ a^2=\dot{x}^2=\text{const},
\end{equation}
and evaluate their action which gives the exponential dependence of
the pair production rate. The sub-leading prefactor is given by
quadratic fluctuations around such solutions. For simple fields,
$iF_{\mu\nu}$ is real (the euclidean potential is purely imaginary)
and~\eqref{eq:instanton-eom} can often be restricted to a 2D-plane,
sometimes even solved analytically~\cite{Dunne2005, Dunne2006a}. In
slightly more complicated fields, solutions can be found using a
shooting method, numerically integrating~\eqref{eq:instanton-eom}
using initial conditions that are varied until the periodicity
condition is met~\cite{Dunne2006}. This is not feasible here, as the
instantons are genuinely 3-dimensional in the parallel polarization
case, and 4-dimensional for other polarizations. Furthermore, they are
not even purely real: we choose the euclidean four potential 
(for parallel polarization)
\begin{equation}
iA_4 = E x_3, \quad iA_3 = 
i \frac{\varepsilon E}{\Omega} \sin\left(\Omega (x_1 - i x_4)\right).
\end{equation}
Without the $x_1$-dependence, this would give real instanton equations
(as considered in~\cite{Prefactor,Pulse-shape}), but in this case real
and imaginary parts mix.

To robustly find instantons and evaluate both the exponent and the
prefactor in such fields, we employ a method that will be discussed in
detail elsewhere~\cite{Discrete}, and only provide the basic ideas
here. Instead of a numerical integration of~\eqref{eq:instanton-eom}
(arising in a saddle point approximation of the continuous path
integral), we discretize the paths in~\eqref{eq:WL-path-integral} into
$N$ points from the beginning and perform the saddle point method on
the resulting $N\times d$-dimensional 
integral, where $d$ is the number of space-time dimensions. 

The equations of
motion~\eqref{eq:instanton-eom} are then replaced by a system of
$N\times d$ nonlinear equations in $N\times d$ unknowns, which can be
solved efficiently using a Newton-Raphson scheme, provided we choose a
sufficiently close initial guess. In our case, the $\gamma\to 0$ limit
corresponds to a static, homogeneous field so we can start with the
known circular instanton, solve for the instanton at a small, finite
$\gamma$ and use that as initial guess for the next value. This
process is called \emph{natural parameter
  continuation}~\cite{Rheinboldt2000}, and is essentially what was
used in~\cite{Gould2017}. We can improve on this using a more
sophisticated predictor-corrector algorithm, also detailed
in~\cite{Discrete}.

Having found instantons for different values of the Keldysh parameter $\gamma$
we can evaluate the instanton action to obtain the leading exponential
contribution to the effective action. We can also evaluate the
prefactor, which is just given by the inverse square root of the
Hessian matrix $\mat{H}$. We do need to regularize zero modes arising
in the integral, due to reparametrization and translational
invariance. We deal with them using the Faddeev-Popov
method~\cite{Faddeev1967,Gordon2015}, exponentiating the Dirac delta
function, which modifies the Hessian to remove zero eigenvalues
(details, again, in~\cite{Discrete}). The final semiclassical result
is then
\begin{multline}
  \label{eq:discrete}
  \frac{\Gamma}{V_{N_0} m^{N_0}} \approx \\
  \left(\frac{E}{E_\mathrm{S}}\right)^{\frac{N_0}{2}}
  \sqrt{\frac{2\pi}{a^\mathrm{cl}}}
  \left(\frac{N}{a^\mathrm{cl}}\right)^{\frac{Nd}{2}}
  \frac{\Phi[x^\mathrm{cl}]
    \ e^{-\frac{E_\mathrm{S}}{E}\mathcal{A}[x^\mathrm{cl}]}}{\sqrt{\det{\mat{H}[x^\mathrm{cl}]}}},
\end{multline}
where $N_0$ is the number of invariant directions, $V_{N_0}$ the
corresponding volume factor, $x^\mathrm{cl}$ the discrete instanton
(the collection of points) and $a^\mathrm{cl}$ its velocity. The
trajectory and $a$ are made dimensionless by rescaling with $m/qE$.

The expression~\eqref{eq:discrete} has the advantage that it is
applicable for every electromagnetic background field, yielding the
correct prefactor including spin effects without having to compare
limiting cases to determine normalization factors. Also the instantons
are independent of the field strength, so as soon as they are
computed, we can evaluate~\eqref{eq:discrete} for many different
values of $E/E_\mathrm{S}$.

\begin{figure}
  \includegraphics[width=\linewidth]{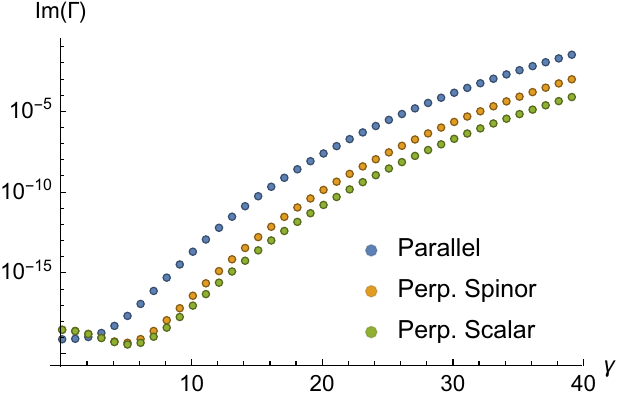}
  \caption{(Color online)
Pair production probabilities for different values of the
    combined Keldysh parameter $\gamma$, strong field strength
    $E=E_\mathrm{S}/30$ (corresponding to a laser intensity of
    $I\approx \SI{5e26}{\watt\per\cm^2}$) and weak field
    $\varepsilon = 10^{-2}$. We have assumed a spacetime volume of
    $\SI{1}{\micro\meter^4}$. As shown before, the case of parallel
    polarization yields stronger enhancement. Further, the spin factor
    does not contribute in the parallel case, while it enhances pair
    production in the perpendicular case.
The values $\Omega=500~\rm keV$ or $\Omega=1~\rm MeV$ considered in 
\cite{Catalysis}, for example, would correspond to $\gamma=30$ or  
$\gamma=60$, respectively, and hence result in a drastic enhancement.
Unfortunately, however, these values are probably outside the
range of near future XFELs.
Lower values such as 25~keV~\cite{EuropeanXFEL} correspond to $\gamma=3/2$ 
(when $E=E_\mathrm{S}/30$) and are thus not sufficient for 
an exponential enhancement. 
On the other hand, for lower field strengths such as $E=E_\mathrm{S}/100$,
the same frequency of 25~keV would correspond to $\gamma=5$ where we start 
to see exponential enhancement. 
However, the total probabilities for $E=E_\mathrm{S}/100$ are much lower 
and thus very hard to detect. 
}
\label{fig:fullPP}
\end{figure}

Fig.~\ref{fig:fullPP} shows evaluations of~\eqref{eq:discrete} for
both parallel and perpendicular polarization, in the perpendicular
case with and without including the spin factor. While there is no
spin dependence in the parallel case, the spin factor further enhances
pair production for perpendicular polarization. The non-monotonic
behavior for small $\gamma$ in the perpendicular case is probably a
sign that the saddle point approximation breaks down, as the instanton
is not confined strongly enough in the $x_4$-direction. This has
already been verified in the purely time dependent case
in~\cite{Prefactor}, where a comparison with the 
(numerical) solution of the Riccati equation
could be made. For the volume factor we have assumed that the strong
field ranges over a four-volume of $\SI{1}{\micro\meter^4}$, which is
completely covered by the plane wave. It does not matter how much
further the plane wave actually extends,  
as it cannot produce any pairs without the strong field. 
We thus hold e.g.\ 
$x_1(0)$, $x_2(0)$ and $x_3(0)$ or their center of mass fixed, 
giving the three-volume 
$V_3$ in~\eqref{eq:discrete} and sum over the instantons located at each 
maximum of the wave giving a factor of $N_\text{inst} = T \Omega / 2\pi$.

%%%%%%%%%%%%%%%%%%%%%%%%%%%%%%%%%%%%%%%%%%%%%%%%%%%%%%%%%%%%%%%%%%%%%%%%%%%%%%%
\section{Conclusions}
%%%%%%%%%%%%%%%%%%%%%%%%%%%%%%%%%%%%%%%%%%%%%%%%%%%%%%%%%%%%%%%%%%%%%%%%%%%%%%%

As an example for the dynamically assisted Sauter-Schwinger effect, we studied 
electron-positron pair creation due to a strong and constant electric field $E$
superimposed by a weaker transversal plane wave with frequency $\Omega$ and 
field strength $\varepsilon E$.  
In analogy to other examples, we found an exponential enhancement of the 
pair-creation probability if the combined Keldysh parameter 
$\gamma=m\Omega/(qE)$ exceeds a threshold value $\gamma_{\rm crit}$ which scales 
in the same way $\gamma_{\rm crit}\sim|\ln\varepsilon|$ as for a purely 
time-dependent sinusoidal field. 
However, the exponential enhancement above threshold $\gamma>\gamma_{\rm crit}$ 
is reduced in comparison to a purely time-dependent sinusoidal field due to 
the effective mass $m_\perp\geq m$ stemming from momentum conservation. 

In order to treat this genuinely space-time dependent field, we employed an 
analytical approach based on a perturbative expansion into powers 
$\varepsilon^N$ of the weaker plane wave (while taking into account the strong 
field $E$ 
non-perturbatively) and compared the results to a 
fully non-perturbative numerical method based on the world-line instanton 
technique. 
Taking into account that the former perturbative approach can yield the 
dominant contribution at a relatively high order $N_*$, we find a good 
agreement with the latter numerical method. 

Note that the world-line instanton action ${\cal A}_{\rm inst}$, 
which yields the exponent in the pair-creation probability 
$P_{e^+e^-}\sim\exp\{-{\cal A}_{\rm inst}\}$,  
depends on polarization and propagation direction.
We find that perpendicular incidence with parallel 
polarization~\eqref{profile} yields maximum enhancement, see also 
\cite{Catalysis}. 
Furthermore, the pre-factor in front of the exponential contains the volume 
scaling. 
Since the plane wave is supposed to be filling the whole volume, this 
pre-factor scales with $L^4$ instead of $L$ as for a single photon, 
cf.~\cite{Catalysis}.

\acknowledgments{G.~Torgrimsson is supported by the Alexander von Humboldt foundation.}

\appendix

%%%%%%%%%%%%%%%%%%%%%%%%%%%%%%%%%%%%%%%%%%%%%%%%%%%%%%%%%%%%%%%%%%%%%%%%%%%%%%%
\section{World-line Formalism}\label{Appendix}
%%%%%%%%%%%%%%%%%%%%%%%%%%%%%%%%%%%%%%%%%%%%%%%%%%%%%%%%%%%%%%%%%%%%%%%%%%%%%%%

Here in the appendix, we use for convenience units with $m=1$ and absorb the charge into the definition of the field strength, i.e. $qE\to E$.

Our starting point is again the worldline representation of the effective action. The spin factor can either be expressed in terms of a path-ordered exponential as in Sec.~\ref{World-line instantons} or as a path integral over an anti-commuting Grassmann variable, $\psi_\mu(\tau)$, with anti-symmetric boundary conditions, $\psi(1)=-\psi(0)$,     
\be\label{GrassmannRepresent}
\begin{split}
\Gamma=2&\int_0^\infty\frac{\ud T}{T}e^{-i\frac{T}{2}}\oint\mathcal{D}x\int\frac{\mathcal{D}\psi}{4} \\
&\exp\left(-i\int_0^1\!\ud\tau\;\frac{\dot{x}^2}{2T}+A\dot{x}-\frac{i}{2}\psi\dot{\psi}-\frac{i}{2}\psi TF\psi\right) \;,
\end{split}
\ee
where $F_{\mu\nu}=\partial_\mu A_\nu-\partial_\nu A_\mu$. In this appendix we consider the superposition of a strong, constant field, $E$, and a plane wave with an arbitrary field shape, given by the potential $A_0=0$ and
$\f{A}(t,\f{r})=Et\f{e}_z+\varepsilon E\f{e}_P\eta(nx)$, where the wave and polarization vectors satisfy $n_\mu=(1,\f{n})$, $\f{n}^2=1$, $\f{e}_P^2=1$, $\f{n}\!\cdot\!\f{e}_P=0$, $nx=t+\f{n}\!\cdot\!\f{r}$, and where $f(nx)=\eta'(nx)$ is, at this point, an arbitrary function; $\eta(nx)=\sin(\Omega nx)/\Omega$ gives the field considered in the main text. With $\varepsilon\ll1$, we expand $\Gamma=\sum_{N=0}^\infty\varepsilon^N\Gamma_N$ and express the weak field in terms of its Fourier transform $\tilde f(\omega)$. 
The center of mass part of the worldline path integral gives an ``energy conserving'' delta function
\be\label{deltaofomegasum}
\int\ud^4x_{\rm cm}\exp\left\{-i\sum_{i=1}^N k_i x_{\rm cm}\right\}=V_32\pi\delta\left(\sum_{i=1}^N \omega_i\right) \;,
\ee 
where $k_{i,\mu}=\omega_i n_\mu$ and $\omega_i$ are the Fourier frequencies corresponding to the $N$ factors of $\tilde{f}$. 	
The rest of the path integral is Gaussian and can be performed as in~\cite{Schubert:2001he,Schubert:2000yt,SchubertLectureNotes} for $N$-photon amplitudes in constant fields. 
This involves the worldline Green's functions $\mathcal{G}_B$ and $\mathcal{G}_F$ for the $x$ and $\psi$ path integrals, respectively. For the exponential part of the probability we only need $\mathcal{G}_B$, which is given by
\be\label{worldlineGreensB}
\begin{split}
\mathcal{G}^B_{\mu\nu}(\tau)
=&-\frac{i}{2E}s\left(2\left[|\tau|-\tau^2\right]-\frac{1}{3}\right)g^\LCperp_{\mu\nu} \\
&-\frac{i}{2E}\left(\frac{\cos[s(1-2|\tau|)]}{\sin s}-\frac{1}{s}\right)g^{\scriptscriptstyle\parallel}_{\mu\nu} \\
&+\frac{\epsilon(\tau)}{2E}\left(\frac{\sin[s(1-2|\tau|)]}{\sin s}-(1-2|\tau|)\right)\hat{F}_{\mu\nu} \;,
\end{split}
\ee
where $s=iET/2$ and where the vector structure is determined by the direction of the strong field, i.e. 
$g^{\scriptscriptstyle\parallel}_{\mu\nu}
=\delta_\mu^0\delta_\nu^0-\delta_\mu^3\delta_\nu^3$, $g_{\mu\nu}^\LCperp=-\delta_\mu^1\delta_\nu^1-\delta_\mu^2\delta_\nu^2$, and $\hat{F}_{\mu\nu}=\delta_\mu^0\delta_\nu^3-
\delta_\mu^3\delta_\nu^0$. 
This Green's function is the Minkowski version of the corresponding Euclidean Green's function, which can be found in~\cite{Schubert:2000yt,Schubert:2001he}.
In terms of this Green's function, we find that the dominant contribution to the $N$-th order is given by 
\be\label{PNGreensB}
\begin{split}
\varepsilon^N &P_N=\text{Im}\int\prod_{i=1}^N\ud\omega_i \tilde{f}(\omega_i)\delta\left[\sum_{i=1}^N\omega_i\right]\int_0^\infty\!\ud s\int_0^1\prod_{i=1}^N\ud\tau_i \\
&\dots\exp\left(-\frac{s}{E}-\frac{i}{2}\sum_{i,j=1}^N k_i[\mathcal{G}_B(\tau_i-\tau_j)
-\mathcal{G}_B(0)]k_j\right) \;,
\end{split}
\ee
where the ellipses stand for sub-leading prefactor terms. The last term in~\eqref{worldlineGreensB} does not contribute, because all $k_i$ are parallel and $k_i\hat{F}k_j=0$.  
The two terms proportional to $g^\LCperp$ and $g^{\scriptscriptstyle\parallel}$ both lead to terms proportional to $n_\LCperp^2\omega_i\omega_j=-k_i g^\LCperp k_j$, so, $n_\LCperp\omega_i$ gives an effective Fourier frequency. Since $|n_\LCperp\omega_i|\leq|\omega_i|$, the exponential is therefore maximized by plane waves travelling perpendicular to the strong field, i.e. for $n_z=0$.
It follows from the delta function~\eqref{deltaofomegasum} that we necessarily have both positive and negative frequencies. We label the frequencies such that $\omega_i>0$ for $i=1,...,J$ and $\omega_i<0$ for $i=J+1,...,N$. Consider each term in the sum in the exponent of~\eqref{PNGreensB} separately. The term proportional to $\omega_i\omega_j$ is maximized by $|\tau_i-\tau_j|=0,1$ for $\omega_i\omega_j>0$ and by $|\tau_i-\tau_j|=1/2$ for $\omega_i\omega_j<0$.  
Similar to the saddle point method, we obtain the dominant exponential contribution by substituting these ``maximizing'' values of $\tau_i$ into~\eqref{PNGreensB}. 
This gives the following exponential for the $s$-integral
\be\label{sExpPlaneWaves}
\exp\left\{-\frac{2m_\LCperp^2}{E}\left(\frac{s}{2}-\Sigma^2\tan\frac{s}{2}\right)\right\} \;,
\ee
where we have defined
\be\label{pperpPWdef}
\Sigma:=\frac{p_\LCperp}{m_\LCperp} \qquad
p_\LCperp:=\frac{1}{2}\sum_{i=1}^Jk_{i\LCperp} \qquad m_\LCperp:=\sqrt{1+p_\LCperp^2}  \;.
\ee
With the saddle point given by $s=2\arccos\Sigma$, we find the general result
\be
\label{SigmaExpPlaneWaves}
\begin{split}
\varepsilon^N P_N\sim&\int\prod_{i=1}^{N}\ud\omega_i\tilde{f}(\omega_i)\delta\left(\sum_{i=1}^N\omega_i\right)\dots \\
&\exp\left\{-\frac{2m_\LCperp^2}{E}\left(\arccos\Sigma-\Sigma\sqrt{1-\Sigma^2}\right)\right\} \\
=&\int\prod_{i=1}^{N}\ud\omega_i\tilde{f}(\omega_i)\delta\left(\sum_{i=1}^N\omega_i\right)\dots \\
&\exp\left\{-\frac{2}{E}\left(-p_\LCperp+m_\LCperp^2\arctan\frac{1}{p_\LCperp}\right)\right\} \;.
\end{split}
\ee
With only one photon, $N=2$, we have $p_\LCperp=k_\LCperp/2$ and the exponential in~\eqref{SigmaExpPlaneWaves} reduces to that in eq.~5 in~\cite{Catalysis}, as expected. Note that the exponential in~\eqref{SigmaExpPlaneWaves} has the same functional dependence of $\Sigma$ as for the longitudinal, purely time-dependent fields we considered in~\cite{non-perturbative-versus-perturbative}, see eq.~3.4 in~\cite{non-perturbative-versus-perturbative}. Comparing $\Sigma$ in~\eqref{pperpPWdef} with the corresponding quantity in eq.~3.3 in~\cite{non-perturbative-versus-perturbative}, we see that the main difference in going from the purely time-dependent fields in~\cite{non-perturbative-versus-perturbative} to the plane waves considered here is the appearance of a heavy effective mass, i.e. $m\to m_\LCperp>m$ (c.f.~\cite{Doubly-assisted}), due to the spatial components of the wave vector. This means that plane waves will in general lead to less exponential enhancement than a purely time-dependent weak field.

\section{Dominant Order}

In this appendix we will present two methods for obtaining estimates of the dominant order, $N_*$, in the instanton formalism. These methods allow us to confirm the dominant order found using the approach in Appendix~\ref{Appendix}.    

\begin{figure}
\includegraphics[width=\linewidth]{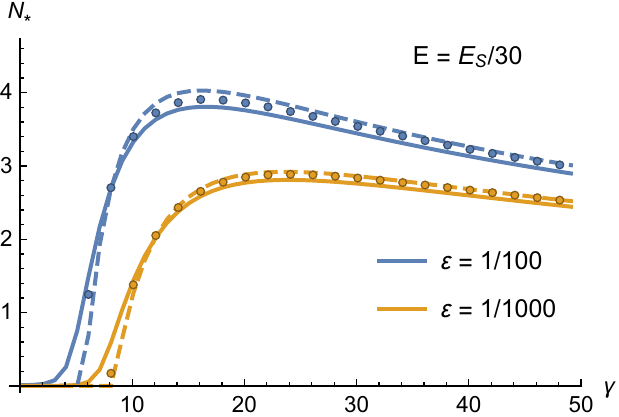}
\includegraphics[width=\linewidth]{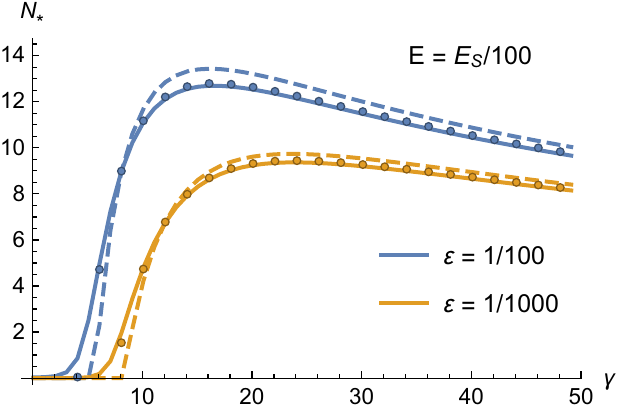}
\caption{These plots show our three estimates of the dominant order for perpendicular polarization. The dashed lines are obtained from equation~\eqref{TranscendentalSaddle}, the solid lines show~\eqref{NdomInstanton}, and the circles show~\eqref{logDerDef}, in which $P_{e^\LCp e^\LCm}$ is the total probability for spinor QED, including the prefactor. The third approach gives negative $N_*$ below the threshold, but that is just another sign (c.f.~Fig.~\ref{fig:fullPP}) that the instanton prediction of the prefactor breaks down in that regime~\cite{NegativeNdom}.}
\label{logDerPlots}
\end{figure}

Our starting point is the worldline representation of the effective action~\eqref{GrassmannRepresent}. In the first method we focus on the scalar part of~\eqref{GrassmannRepresent}, i.e. the part without Grassmann variables or Dirac matrices; we will show below using the second method that the spin factor does not significantly affect $N_*$. Let $a_\mu
\propto\varepsilon$ be the the weak field. For the fields we focus on in this paper, $a_\mu$ is a plane wave, but the methods we present here work also for more general field shapes. We expand the exponent in the weak field,
\be\label{startNdomInstanton}
\exp\left\{-i\int_0^1 a\dot{x}\right\}=\sum_{N=0}^\infty\frac{1}{N!}\left(-i\int_0^1 a\dot{x}\right)^N \;,
\ee
and obtain an estimate of the dominant order from the ``saddle point'' of the sum over $N$. Assuming that the dominant order is ``large'', we use Stirling's approximation $\ln N!\approx N(\ln N-1)$ and find
\be\label{NdomInstanton}
N_*=-i\int_0^1 a\dot{x} \;.
\ee
At this order we recover the original exponent, i.e.
\be \label{originalExponentAgain}
\frac{1}{N_*!}\left(-i\int_0^1 a\dot{x}\right)^{N_*}\approx\exp\left\{-i\int_0^1 a\dot{x}\right\} \;.
\ee 
The instantons and the resulting exponential part of the probability will therefore be exactly the same as before, i.e. as without the additional steps~\eqref{startNdomInstanton} to~\eqref{originalExponentAgain}. The point is that substituting the instantons into~\eqref{NdomInstanton} gives us a simple estimate of the dominant order in the instanton formalism. Note that, while we assume that $a_\mu$ is weak, the integral of $a_\mu$ in~\eqref{NdomInstanton} gives $N_*$, which is supposed to be large. So, in~\eqref{startNdomInstanton} we expand the exponent in a parameter which is actually large. That is of course not a problem as it only means that we have to sum up many terms (the Taylor series for the exponential has infinite radius of convergence). In fact, we want this expansion parameter to be large because in regimes where it is small, the dominant contribution comes from $N=0$ and then there is no significant enhancement of the probability.  
As shown in Fig.~\ref{logDerPlots}, the instanton estimate~\eqref{NdomInstanton} agrees with the previous estimate based on~\eqref{TranscendentalSaddle}. So, \eqref{NdomInstanton} seems to give a good estimate of the dominant order in dynamical assistance.          

A more direct way of estimating the dominant order is to calculate the logarithmic derivative of the probability with respect to $\varepsilon$, i.e.
\be\label{logDerDef}
N_*=\frac{\ud\log P_{e^\LCp e^\LCm}}{\ud\log\varepsilon} \;.
\ee
This expression is motivated by the fact that in a regime where $P_{e^\LCp e^\LCm}\sim\varepsilon^{N_0}$, \eqref{logDerDef} gives $N_*=N_0$. 
In Fig.~\ref{logDerPlots} we evaluate~\eqref{logDerDef} within the instanton formalism, and show that~\eqref{logDerDef} agrees quite well with the simpler estimate of~\eqref{NdomInstanton}. One advantage of~\eqref{logDerDef} is that it is general and does not depend on how we calculate $P_{e^\LCp e^\LCm}$. For example, for purely time dependent fields, like the ones we studied in~\cite{non-perturbative-versus-perturbative}, one can obtain the exact probability by solving the Riccati equation numerically, and then~\eqref{logDerDef} gives an exact description of how the probability depends on~$\varepsilon$.      

All the plots in Fig~\ref{logDerPlots} show qualitatively the same behavior as a function of $\gamma$: Below the threshold $N_*\approx0$ (for $\varepsilon$ sufficiently small compared to $E$)~\cite{NegativeNdom}, where the probability is given by Schwinger's constant field result. After the threshold, $N_*$ quickly reaches a maximum and then slowly decreases as $\gamma\to\infty$. So, pair creation actually becomes less ``multi-photon'' as $\gamma$ increases beyond the maximum. 
As Fig~\ref{logDerPlots} shows, the maximum dominant order can be quite large. However, it is not large for all relevant parameters. For e.g. $E=1/30$ and $\varepsilon=1/100$, the maximum dominant order is only $N_*\sim4$, which suggests that it might be feasible in this case to actually calculate also the pre-exponential contributions to all important orders.

%%%%%%%%%%%%%%%%%%%%%%%%%%%%%%%%%%%%%%%%%%%%%%%%%%%%%%%%%%%%%%%%%%%%%%%%%%%%%%%
%%%%%%%%%%%%%%%%%%%%%%%%%%%%%%%%%%%%%%%%%%%%%%%%%%%%%%%%%%%%%%%%%%%%%%%%%%%%%%%


\begin{thebibliography}{99}
%%%%%%%%%%%%%%%%%%%%%%%%%%%%%%%%%%%%%%%%%%%%%%%%%%%%%%%%%%%%%%%%%%%%%%%%%%%%%%%
%%%%%%%%%%%%%%%%%%%%%%%%%%%%%%%%%%%%%%%%%%%%%%%%%%%%%%%%%%%%%%%%%%%%%%%%%%%%%%%

\bibitem{Schroedinger}
E.~Schr\"odinger, 
{\em The proper vibrations of the expanding Universe}, 
Physica {\bf 6}, 899 (1939). 

\bibitem{Hawking74}
S.~W.~Hawking,
{\em Black Hole Explosions},
Nature {\bf 248}, 30 (1974).

\bibitem{Hawking75}
S.~W.~Hawking,
{\em Particle Creation by Black Holes},
Commun.\ Math.\ Phys.\ {\bf 43}, 199 (1975).

%\cite{Sauter1931}
\bibitem{Sauter1931} 
F.~Sauter,
{\em \"Uber das Verhalten eines Elektrons im homogenen elektrischen Feld 
nach der relativistischen Theorie Diracs},
Z.\ Phys.\  {\bf 69}, 742 (1931).
%  doi:10.1007/BF01339461
  %%CITATION = doi:10.1007/BF01339461;%%
  %443 citations counted in INSPIRE as of 20 Nov 2017

%\cite{Sauter1932}
\bibitem{Sauter1932} 
F.~Sauter,
{\em Zum ``Kleinschen Paradoxon''},
Z.\ Phys.\  {\bf 73}, 547 (1932).
%  doi:10.1007/BF01349862
  %%CITATION = doi:10.1007/BF01349862;%%
  %86 citations counted in INSPIRE as of 20 Nov 2017

%\cite{Heisenberg1936}
\bibitem{Heisenberg1936} 
W.~Heisenberg and H.~Euler,
{\em Consequences of Dirac's theory of positrons}, 
Z.\ Phys.\  {\bf 98}, 714 (1936).
%  doi:10.1007/BF01343663
%  [physics/0605038].
  %%CITATION = doi:10.1007/BF01343663;%%
  %1379 citations counted in INSPIRE as of 20 Nov 2017

%\cite{Schwinger}
\bibitem{Schwinger} 
J.~S.~Schwinger,
{\em On gauge invariance and vacuum polarization},
  Phys.\ Rev.\  {\bf 82}, 664 (1951).
%  doi:10.1103/PhysRev.82.664
  %%CITATION = doi:10.1103/PhysRev.82.664;%%
  %4341 citations counted in INSPIRE as of 20 Nov 2017

%\cite{Dynamically-assisted}
\bibitem{Dynamically-assisted} 
R.~Sch\"utzhold, H.~Gies and G.~Dunne,
{\em Dynamically assisted Schwinger mechanism},
  Phys.\ Rev.\ Lett.\  {\bf 101}, 130404 (2008). 
%  doi:10.1103/PhysRevLett.101.130404
%  [arXiv:0807.0754 [hep-th]].
  %%CITATION = doi:10.1103/PhysRevLett.101.130404;%%
  %160 citations counted in INSPIRE as of 21 Nov 2017

\bibitem{footnote} 
As an exception, it was shown in \cite{Pulse-shape}
that adding a standing 
(plane) wave $\propto\cos(kx)\cos(\omega t)$ to a constant field $E$ 
yields the same enhancement exponent as the corresponding purely 
time-dependent field.
%
Furthermore, in \cite{inhomogeneous} a spatially varying strong field plus 
a time-dependent weaker field is considered, which leads to a non-trivial 
interplay between the spatial and temporal dependences.  

%\cite{Pulse-shape}
\bibitem{Pulse-shape} 
M.~F.~Linder, C.~Schneider, J.~Sicking, N.~Szpak and R.~Sch\"utzhold,
{\em Pulse shape dependence in the dynamically assisted Sauter-Schwinger 
effect},
  Phys.\ Rev.\ D {\bf 92}, no. 8, 085009 (2015).
%  doi:10.1103/PhysRevD.92.085009
%  [arXiv:1505.05685 [hep-th]].
  %%CITATION = doi:10.1103/PhysRevD.92.085009;%%
  %21 citations counted in INSPIRE as of 21 Nov 2017

%\cite{inhomogeneous}
\bibitem{inhomogeneous} 
C.~Schneider and R.~Sch\"utzhold,
{\em Dynamically assisted Sauter-Schwinger effect in inhomogeneous 
electric fields},
JHEP {\bf 1602}, 164 (2016). 
%  doi:10.1007/JHEP02(2016)164
%  [arXiv:1407.3584 [hep-th]].
  %%CITATION = doi:10.1007/JHEP02(2016)164;%%
  %27 citations counted in INSPIRE as of 21 Nov 2017
    
%\cite{Narozhnyi:1976zs}
\bibitem{Narozhnyi:1976zs}
  N.~B.~Narozhnyi and A.~I.~Nikishov,
  {\em Solutions of the Klein-Gordon and Dirac Equations for a Particle in a Constant Electric Field and an Electromagnetic Wave Parallel to It},
  Teor.\ Mat.\ Fiz.\  {\bf 26} 16 (1976).
  %doi:10.1007/BF01038251
  %%CITATION = doi:10.1007/BF01038251;%%
  %17 citations counted in INSPIRE as of 13 Jun 2017  
  
%\cite{Fradkin:1991zq}
\bibitem{Fradkin:1991zq}
  E.~S.~Fradkin, D.~M.~Gitman and S.~M.~Shvartsman,
  {\em Quantum electrodynamics with unstable vacuum},
  (Springer series in nuclear and particle physics, Berlin, 1991).
  %30 citations counted in INSPIRE as of 16 Jun 2017  
  
%\cite{Gitman:1996wm}
\bibitem{Gitman:1996wm}
  D.~M.~Gitman and S.~I.~Zlatev,
  {\em Spin factor in path integral representation for Dirac propagator in external fields},
  Phys.\ Rev.\ D {\bf 55} 7701 (1997).
  %doi:10.1103/PhysRevD.55.7701
  %[hep-th/9608179].
  %%CITATION = doi:10.1103/PhysRevD.55.7701;%%
  %18 citations counted in INSPIRE as of 14 Jun 2017  

%\cite{Tomaras:2000ag}
\bibitem{Tomaras:2000ag}
  T.~N.~Tomaras, N.~C.~Tsamis and R.~P.~Woodard,
  {\em Back reaction in light cone QED},
  Phys.\ Rev.\ D {\bf 62} 125005 (2000).
  %doi:10.1103/PhysRevD.62.125005
  %[hep-ph/0007166].
  %%CITATION = doi:10.1103/PhysRevD.62.125005;%%
  %38 citations counted in INSPIRE as of 15 Nov 2017  

%\cite{Tomaras:2001vs}
\bibitem{Tomaras:2001vs}
  T.~N.~Tomaras, N.~C.~Tsamis and R.~P.~Woodard,
  {\em Pair creation and axial anomaly in light cone QED(2)},
  JHEP {\bf 0111} 008 (2001).
  %doi:10.1088/1126-6708/2001/11/008
  %[hep-th/0108090].
  %%CITATION = doi:10.1088/1126-6708/2001/11/008;%%
  %28 citations counted in INSPIRE as of 15 Nov 2017  
    
%\cite{Ilderton:2014mla}
\bibitem{Ilderton:2014mla}
  A.~Ilderton,
  {\em Localisation in worldline pair production and lightfront zero-modes},
  JHEP {\bf 1409} 166 (2014).
  %doi:10.1007/JHEP09(2014)166
  %[arXiv:1406.1513 [hep-th]].
  %%CITATION = doi:10.1007/JHEP09(2014)166;%%
  %18 citations counted in INSPIRE as of 13 Jun 2017  

%\cite{Ilderton:2015qda}
\bibitem{Ilderton:2015qda}
  A.~Ilderton, G.~Torgrimsson and J.~W{\aa}rdh,
  {\em Nonperturbative pair production in interpolating fields},
  Phys.\ Rev.\ D {\bf 92} no.6, 065001 (2015).
  %doi:10.1103/PhysRevD.92.065001
  %[arXiv:1506.09186 [hep-th]].
  %%CITATION = doi:10.1103/PhysRevD.92.065001;%%
  %22 citations counted in INSPIRE as of 30 Nov 2017

%\cite{Catalysis}
\bibitem{Catalysis} 
G.~V.~Dunne, H.~Gies and R.~Sch\"utzhold,
{\em Catalysis of Schwinger Vacuum Pair Production},
Phys.\ Rev.\ D {\bf 80}, 111301 (2009).
%  doi:10.1103/PhysRevD.80.111301
%  [arXiv:0908.0948 [hep-ph]].
  %%CITATION = doi:10.1103/PhysRevD.80.111301;%%
  %111 citations counted in INSPIRE as of 21 Nov 2017
  
%\cite{Schubert:2001he}
\bibitem{Schubert:2001he}
  C.~Schubert,
  {\em Perturbative quantum field theory in the string inspired formalism},
  Phys.\ Rept.\ {\bf 355} 73 (2001).
  %doi:10.1016/S0370-1573(01)00013-8
  %[hep-th/0101036].
  %%CITATION = doi:10.1016/S0370-1573(01)00013-8;%%
  %215 citations counted in INSPIRE as of 05 May 2016

\bibitem{SchubertLectureNotes}
C.~Schubert,
{\em Lectures on the Worldline Formalism},
School on Spinning Particles in Quantum Field Theory:
Worldline Formalism, Higher Spins, and Conformal Geometry

\bibitem{example}
For example, in the limit $\Omega\to0$ we have a constant field with field 
strength $E(1+\varepsilon)$, and by expanding Schwinger's result one finds 
that all orders in $\varepsilon$ are nonzero.
  
\bibitem{Affleck1982}
I.~K.~Affleck, O.~Alvarez, and N.~S.~Manton, {\em Pair production
at strong coupling in weak external fields}, Nucl. Phys. B
{\bf 197}(3), 509–519 (1982).

\bibitem{Dunne2005}
G.~V.~Dunne and C.~Schubert, {\em Worldline instantons and pair
production in inhomogenous fields}, Phys. Rev. D {\bf 72}(10),
105004 (2005).

\bibitem{Dunne2006a}
G.~V.~Dunne, Q.~Wang, H.~Gies and C.~Schubert, {\em Worldline
instantons and the fluctuation prefactor}, Phys. Rev. D {\bf 73}(6),
065028 (2006).

\bibitem{Dunne2006}
G.~V.~Dunne and Q.~Wang, {\em Multidimensional worldline
instantons}, Phys. Rev. D {\bf 74}(6),
065015 (2006).

\bibitem{Discrete}
C.~Schneider~et~al.,
%G.~Torgrimsson and R.~Sch\"utzhold, 
{\em Discrete worldline instantons}, in preparation.

\bibitem{Rheinboldt2000}
W.~Rheinboldt, {\em Numerical continuation methods: a
perspective}, Journal of computational and applied mathematics
{\bf 124}, 229–244 (2000).

\bibitem{Gould2017}
O.~Gould and A.~Rajantie, {\em Thermal Schwinger pair production
  at arbitrary coupling}, Phys. Rev. D {\bf 96}(7) (2017).

\bibitem{Faddeev1967}
L.~D.~Faddeev and V.~N.~Popov, {\em Feynman diagrams for the
  Yang-Mills field}, Phys. Lett. {\bf 25B}(1),
  29–30 (1967).

\bibitem{Gordon2015}
J.~Gordon and G.~W.~Semenoff, {\em World-line instantons and the 
Schwinger effect as a Wentzel-Kramers-Brillouin exact path integral}, 
Journal of Mathematical Physics {\bf 56}(2) (2015).

%\cite{non-perturbative-versus-perturbative}
\bibitem{non-perturbative-versus-perturbative} 
G.~Torgrimsson, C.~Schneider, J.~Oertel and R.~Sch\"utzhold,
{\em Dynamically assisted Sauter-Schwinger effect --  
non-perturbative versus perturbative aspects},
  JHEP {\bf 1706}, 043 (2017).
%  doi:10.1007/JHEP06(2017)043
%  [arXiv:1703.09203 [hep-th]].
  %%CITATION = doi:10.1007/JHEP06(2017)043;%%
  %6 citations counted in INSPIRE as of 21 Nov 2017

%\cite{Doubly-assisted}
\bibitem{Doubly-assisted} 
G.~Torgrimsson, J.~Oertel and R.~Sch\"utzhold,
{\em Doubly assisted Sauter-Schwinger effect},
  Phys.\ Rev.\ D {\bf 94}, no. 6, 065035 (2016).
%  doi:10.1103/PhysRevD.94.065035
%  [arXiv:1607.02448 [hep-th]].
  %%CITATION = doi:10.1103/PhysRevD.94.065035;%%
  %1 citations counted in INSPIRE as of 21 Nov 2017
  
\bibitem{EuropeanXFEL}
European XFEL,
\url{www.xfel.eu}
  

%\cite{Prefactor}
\bibitem{Prefactor} 
C.~Schneider and R.~Sch\"utzhold,
{\em Prefactor in the dynamically assisted Sauter-Schwinger effect},
Phys.\ Rev.\ D {\bf 94}, no. 8, 085015 (2016).
%  doi:10.1103/PhysRevD.94.085015
%  [arXiv:1603.00864 [hep-th]].
  %%CITATION = doi:10.1103/PhysRevD.94.085015;%%
  %5 citations counted in INSPIRE as of 21 Nov 2017


%\cite{Schubert:2000yt}
\bibitem{Schubert:2000yt}
  C.~Schubert,
  {\em Vacuum polarization tensors in constant electromagnetic fields. Part 1.}, 
  Nucl.\ Phys.\ B {\bf 585} 407 (2000).
  %doi:10.1016/S0550-3213(00)00423-5
  %[hep-ph/0001288].
  %%CITATION = doi:10.1016/S0550-3213(00)00423-5;%%
  %49 citations counted in INSPIRE as of 15 Apr 2017  
 
\bibitem{NegativeNdom}
If one estimates the dominant order by calculating $P_{e^\LCp e^\LCm}$ in~\eqref{logDerDef} using the instanton formalism, then one might obtain incorrect results below the threshold, e.g. $N_*\approx-1$. This is because, in that regime, terms in the exponent that were assumed to be large are actually not. One can understand this in the limit $\gamma\to0$ by expanding the (locally) constant field result in $\varepsilon$.        
  
  

\end{thebibliography}
\end{document}